# Unsupervised single-particle deep clustering via statistical manifold learning

Jiayi Wu[1,2], Yong-Bei Ma[2], Charles Congdon[3], Bevin Brett[3], Shuobing Chen[1,2], Yaofang Xu[2,4], Qi Ouyang[1], Youdong Mao[1,2,5]*

[1]State Key Laboratory for Artificial Microstructure and Mesoscopic Physics, Institute of Condense Matter Physics, School of Physics, Center for Quantitative Biology, Peking University, Beijing 100871, China.

[2]Intel Parallel Computing Center for Structural Biology, Dana-Farber Cancer Institute, Boston, MA 02215, USA.

[3]Software and Services Group, Intel Corporation, Santa Clara, CA 95052, USA.

[4]Department of Biophysics, Peking University Health Science Center, Beijing 100191, China.

[5]Department of Microbiology and Immunobiology, Harvard Medical School, Boston, MA 02115, USA.

*Correspondence should be addressed to Y.M. Tel: +1 617 632 4358. Fax: +1 617 632 4338. E-mail address: youdong_mao@dfci.harvard.edu

**Abstract**

**Motivation:** Structural heterogeneity in single-particle cryo-electron microscopy (cryo-EM) data represents a major challenge for high-resolution structure determination. Unsupervised classification may serve as the first step in the assessment of structural heterogeneity. Traditional algorithms for unsupervised classification, such as K-means clustering and maximum likelihood optimization, may classify images into wrong classes with decreasing signal-to-noise-ratio (SNR) in the image data, yet demand increased cost in computation. Overcoming these limitations requires further development on clustering algorithms for high-performance cryo-EM data analysis.

**Results:** Here we introduce a statistical manifold learning algorithm for unsupervised single-particle deep clustering. We show that statistical manifold learning improves classification accuracy by about 40% in the absence of input references for lower SNR data. Applications to several experimental datasets suggest that our deep clustering approach can detect subtle structural difference among classes. Through code optimization over the Intel high-performance computing (HPC) processors, our software implementation can generate thousands of reference-free class averages within several hours from hundreds of thousands of single-particle cryo-EM images, which allows significant improvement in *ab initio* 3D reconstruction resolution and quality. Our approach has been successfully applied in several structural determination projects. We expect that it provides a powerful computational tool in analyzing highly heterogeneous structural data and assisting in computational purification of single-particle datasets for high-resolution reconstruction.

**Availability:** Our software implementation in C++ is available at http://ipccsb.dfci.harvard.edu/rome. The experimental cryo-EM datasets used for benchmarking the algorithm are available at Electron Microscopy Pilot Image Archive, www.ebi.ac.uk/pdbe/emdb/empiar with accession no. EMPIAR-10069.

**Contact:** youdong_mao@dfci.harvard.edu

## 1  Introduction

Single-particle cryo-EM is emerging as a mainstream tool to visualize the three-dimensional (3D) structures of biomolecules in their native functional states at near-atomic resolution (Frank, 2006; Nogales, 2016). Individual biomolecules may sample multiple conformations. Thus a prerequisite for high-resolution structure determination is to obtain a 'pure' dataset that includes particle images in an identical conformational state (Scheres, et al., 2007). To distinguish different conformations in raw particle images, reference-free two-dimensional (2D) classification is commonly practiced as the first step in evaluation of structural heterogeneity (Frank, 2006; Katsevich, et al., 2015; Schwander, et al., 2014; Tagare, et al., 2015), or used as an intermediate step during the alteration of 2D and 3D classifications for *in silico* purification (Chen, et al., 2016; Zhang, et al., 2015). Besides, high-quality reference-free 2D class averages are essential for reliable initial reconstruction and its verification (Frank, 2006; Murray, et al., 2013; Zhang, et al., 2012).

Though several classification schemes were proposed for single-particle analysis (Pascual-Montano, et al., 2001; Singer, et al., 2011; Sorzano, et al., 2010; Yang, et al., 2012; Zhao and Singer, 2014), there are mainly two approaches for unsupervised 2D classification of single-particles implemented in multiple end-user software packages (Baldwin and Penczek, 2007; de la Rosa-Trevin, et al., 2013; Hohn, et al., 2007; Scheres, 2012; Shaikh, et al., 2008; Sorzano, et al., 2004) and widely used in cryo-EM structure determination: (1) K-means clustering following reference-free alignment through cross-correlation (CC) and multivariate statistical analysis (MSA) (Frank, 2006; Van Heel and Frank, 1980; Van Heel and Frank, 1981; van Heel and Stöffler-Meilicke, 1985), and (2) unsupervised maximum-likelihood (ML) or *maximum a posteriori* (MAP) classification (Scheres, 2012; Scheres, et al., 2005; Sigworth, 1998). In the former approach, the classification accuracy is affected by the noise-induced misalignment resulting from false peaks in cross-correlation calculation. Noise may also introduce errors in the distance calculation in K-means clustering. Hence, its performance dramatically reduces as the signal-to-noise ratio (SNR) decreases. By contrast, the ML-based approach explores optimal probability in measuring image similarity, and exhibits robust resistance to the noise-induced misalignment (Sigworth, 1998). However, a prominent drawback lies in that the likelihood matching insufficiently differentiates structural heterogeneity among similar but critically different views. In each of ML-classified group of single-particles one could find a mixture of heterogeneous 2D projection structures with a large variation in likelihood. This effect causes a decrease in the effective number of classes while increasing the cycles of ML optimization. To date, there has been a lack of reference-free classification methods that can efficiently sort out highly heterogeneous single-particles into thousands of homogenous 2D classes while still keeping computation efficient (Schwander, et al., 2014).

Recently, manifold embedding has been applied to study the continuous conformational changes of 80S ribosomes by single-particle cryo-EM (Dashti, et al., 2014). In this approach, a one-dimensional manifold was used to describe a continuous conformational spectrum that can be discerned through a narrow angular aperture. The approach begins with a classification of data into different orientations with respect to a common template, with the assumption that the changes in the structure are relatively small so that a projection-matching based refinement algorithm with a single template can be used. However, this assumption does not necessarily hold for any molecules exhibiting dramatic conformational changes, in which a small angular aperture is not sufficient to distinguish the conformational spectrum. Hence, such an approach is yet to be adapted to a more general condition.

Here we introduce a reference-free deep clustering method based on statistical manifold learning (SML) without any assumptions regarding how the molecules assume their conformations in cryo-EM data (Bishop, et al., 1998; Bishop, et al., 1998; Gorban, et al., 2008). We implemented SML-based clustering in a software package named ROME (Refinement and Optimization based on Machine lEarning), which was optimized against both Intel® Xeon® multi-core processors and Intel® Many-Integrated Core (MIC) Architecture (Jeffers and Reinders, 2013) to enable efficient computation of thousands of reference-free class averages in a highly affordable fashion. The increased number of reference-free class averages prominently improved the quality and resolution of *ab initio* 3D models with angular reconstitution. We further tested our approach with several cryo-EM datasets to demonstrate the advantage of deep clustering in discerning subtle structural differences directly in 2D class averages corresponding to distinct conformations.

## 2 Methods

### 2.1 Mathematical model for statistical manifold learning

Our SML algorithm is derived from generative topographic mapping (GTM) algorithm framework (Bishop, et al., 1998; Bishop, et al., 1998; Gorban, et al., 2008). The goal of GTM is to find a representation for the distribution $p(t)$ of dataset in a $J$-dimensional data space $t = (t_1, \ldots, t_J)$ in terms of $L$-dimensional latent variables $s = (s_1, \ldots, s_L)$. One may consider a non-linear, parametric function $t = A(s; W)$, which maps points $s$ in the latent space into corresponding points $A(s; W)$ in the data space (Bishop, et al., 1998) (Fig. 1a). The mapping is controlled by a set of parameters $W$, which would represent the weights and biases in the case of a feedforward neural network as the mapping. In the situation in which the dimensionality $L$ of the latent space is less than the dimensionality $J$ of the data space, the nonlinear transformation $A(s; W)$ maps the latent space into an $L$-dimensional non-Euclidean manifold $S$ embedded within the data space. Previous studies have established GTM as an alternative to self-organizing maps (Allinson, 2001) (SOM) and that the GTM framework overcomes most limitations in the SOM while introducing no significant disadvantages (Bishop, et al., 1998; Bishop, et al., 1998).

To adapt the general framework of GTM to single-particle cryo-EM data classification problem, we built the contrast transfer function (CTF) into the non-linear mapping function in Fourier or reciprocal space, which paved the way to efficiently correct the aberration effect of the objective lenses in electron microscopy (Frank, 2006). A vector in data space $X_i$ represents the Fourier transform of a particle image. Thus an image vector $X_i$ in Fourier space can be modelled as:

$$T_{-\tau}(X_{ij}) = \text{CTF}_{ij}\left[A(s;W)\right]_j + N_{ij}, \qquad (1)$$

where $[\cdot]_j$ is the $j$-th pixel in an Fourier-transformed image; $T_{-\tau}$ denotes the in-plane image rotation and translation operator; $\tau=(\theta,x,y)$ is the in-plane rotation angle and translation; $X_{ij}$ denotes the intensity of the $j$-th pixel of the $i$-th observed image in Fourier space; $\text{CTF}_{ij}$ denotes the $j$-th component of contrast transfer function for the $i$-th image; $N_{ij}$ denotes the white Gaussian noise and $[A(s; W)]_j$ denotes the $j$-th pixel of the molecular projection in Fourier space for image $X_i$. The latent variables *s* would reflect the inherent degrees of freedom controlling 2D structural differences observed in single-particle images, which arise from either distinct conformational states of the imaged biomolecules or different viewing angles. For simplification purpose, we use $Y_i$ to denote the *i*-th translated and rotated single-particle image in Fourier space; thus, $Y_{ij}=T_{-\tau}(X_{ij})$ corresponds to the *j*-th pixel of the *i*-th translated and rotated image in Fourier space. The geometrical parameters $\tau=(\theta,x,y)$ can be determined through a 2D image alignment procedure, such as the regularized maximum-likelihood method (Scheres, 2012; Scheres, et al., 2005; Sigworth, 1998).

The nonlinear function $A(s; W)$ can be expanded by a set of basis functions $\{\varphi_m\}$ through a weight matrix **W**:

$$[A(s;W)]_j = \sum_{m=1}^{M} \varphi_m(s) W_{mj}. \qquad (2)$$

Here **W** is a $M \times J$ matrix containing the weight and bias parameters. In our algorithmic design, we use a combination of one fixed basis function and many Gaussians basis functions in the form

$$\varphi_m(s) = \begin{cases} \exp\{-\frac{\|s-\mu_m\|^2}{2\sigma^2}\}, m \leq M_{NL} \\ 1, m = M_{NL}+1 \end{cases}. \qquad (3)$$

Here $M_{NL}$ denotes the number of Gaussian basis functions; $\mu_m$ denotes the mean of the Gaussian distribution and $\sigma$ denotes the common variance of the Gaussian distribution.

Although the noise in cryo-EM may take multiple forms, potentially including both white and non-while noises, previous studies have established that the cryo-EM noise can be largely approximated with a normal distribution without obvious detrimental effect in data analysis (Frank, 2006). Thus, in our algorithm the probability density function in data space is chosen as an isotropic Gaussian distribution centered on CTF[$A(s; W)$] in the form

$$p(Y_i|s,W,\beta) = \prod_{j=1}^{J} \left(\frac{\beta_{ij}}{2}\right)^{\frac{1}{2}} \exp\{-\frac{\beta_{ij}}{2}\left(Y_{ij} - \text{CTF}_{ij}\left[A(s;W)\right]_j\right)^2\}, \quad (4)$$

where $\beta$ is the variance matrix of noise, in which $\beta_{ij}$ is the variance of noise for the $j$-th pixel of the $i$-th image.

Since we manage to classify a dataset consisting of $N$ images $\{Y_1, Y_2, …, Y_N\}$ into $K$ classes, this can be translated into the problem of finding $K$ points $\{s_1, s_2, …, s_K\}$ in the latent space that are mapped to $\{Y_1, Y_2, …, Y_N\}$ in the data space through Eq. (1). To allow the problem more tractable, we consider a specific form for the probability distribution $p(s)$ given by a sum of delta functions centered on $K$ nodes of a regular grid in the latent space:

$$p(s) = \frac{1}{K}\sum_{k=1}^{K}\delta(s-s_k). \quad (5)$$

By integrating over $s$-distribution on the manifold, the distribution $p(Y_i|W,\beta)$ in the data space for a given value of $W$ is:

$$p(Y_i|W,\beta) = \int p(Y_i|s,W,\beta)p(s)ds = \frac{1}{K}\sum_{k=1}^{K}p(Y_i|s_k,W,\beta). \quad (6)$$

The joint probability density of observing the $N$ images $\{Y_1, Y_2, …, Y_N\}$ is

$$p(Y_1,…,Y_N|W,\beta) = \prod_{i=1}^{N}p(Y_i|W,\beta).$$ A maximum-likelihood estimate of the model parameters $\Theta=\{W,\beta\}$ can be found by maximizing the logarithmic form of the joint probability, namely, $\hat{\Theta} = \underset{\Theta}{\text{argmax}}\sum_{i=1}^{N}\log p(Y_i|\Theta)$. We would like to consider the prior probability of the weight matrix, in which case a regularized maximum-likelihood estimator, also called the *maximum a*

*posteriori* (MAP) with respect to $\Theta=\{\mathbf{W},\boldsymbol{\beta}\}$ can be used to solve the nonlinear mapping problem. In this case, the model parameters are sought to maximize the MAP:

$$\hat{\Theta}=\operatorname{argmax}[\sum_{i=1}^{N}\log p(Y_i|\Theta)+\log p(\Theta)]. \tag{7}$$

Here we assume an isotropic Gaussian prior distribution over **W**:

$$p(\mathbf{W})=\left(\frac{\alpha}{2\pi}\right)^{\frac{MJ}{2}}\exp(-\frac{\alpha}{2}\sum_{m=1}^{M}\sum_{j=1}^{J}W_{mj}^2). \tag{8}$$

where $\alpha$ is the variance of Gaussian prior over the weight matrix **W.**

## 2.2 Expectation-maximization algorithm in statistical manifold learning

Unsupervised classification problem is ill posed because of a high level of noise, potentially missing orientations and discontinuity between different conformational states in the single-particle cryo-EM data. To enable effective computation, the expectation-maximization (E-M) algorithm can be used to estimate the model parameters $\Theta$ corresponding to an optimal classification. The E-M algorithm alternates between two steps: the expectation step (E-step) and the maximization step (M-step). The inclusion of CTF effect invalidates the original matrix operation formulae in the E-M algorithm to solve the GTM problem (Bishop, et al., 1998). Thus, a new set of mathematical equations were developed for the CTF-embedded E-M algorithm that can properly classify images based on the true 2D structural difference instead of the difference in defocus (see Supplementary Text 1 for mathematical details).

In the E-step, we use the old model parameter $\mathbf{W}^{[n]}$ and $\boldsymbol{\beta}^{[n]}$ to evaluate the posterior probabilities, or responsibilities, $R_{ki}(\Theta^{[n]})$ of each latent variable $s_k$ for every images $Y_i$ using the Bayes' theorem in the form:

$$R_{ki}(\Theta^{[n]})=p(s_k|Y_i,\mathbf{W}^{[n]},\boldsymbol{\beta}^{[n]}) = \frac{p(Y_i|s_k,\mathbf{W}^{[n]},\boldsymbol{\beta}^{[n]})p(s_k)}{\sum_{k'=1}^{K}p(Y_i|s_{k'},\mathbf{W}^{[n]},\boldsymbol{\beta}^{[n]})p(s_{k'})}. \tag{9}$$

In the M-step, we consider maximizing the MAP estimator $Q(\Theta,\Theta^{[n]})$ with respect to the model parameters $\Theta$:

$$Q(\Theta,\Theta^{[n]}) = \sum_{i=1}^{N}\sum_{k=1}^{K}R_{ki}(\Theta^{[n]})\ln p(Y_i|\Theta) + \ln p(\Theta). \tag{10}$$

To this end, the partial derivatives of Eq. (10) with respect to each of the model parameters should equal zero when the MAP is at its maximum. This allows us to estimate the model parameters iteratively. Maximizing Eq. (10) with respect to $\mathbf{W}$ thus gives rise to the following equation, which uses $R_{ki}(\Theta^{[n]}), \beta_{ij}^{[n]}$ to update the weight matrix $\mathbf{W}^{[n+1]}$ in the [$n$+1]-th iteration:

$$\sum_{i=1}^{N}\sum_{k=1}^{K}R_{ki}(\Theta^{[n]})\beta_{ij}^{[n]}\varphi_m(s_k)\mathrm{CTF}_{ij}(Y_{ij} - \mathrm{CTF}_{ij}\sum_{m'=1}^{M}\varphi_{m'}(s_k)W_{m'j}^{[n+1]}) - \alpha W_{mj}^{[n+1]} = 0. \tag{11}$$

Similarly, maximizing Eq. (10) with respect to $\boldsymbol{\beta}$, we obtained the following re-estimation formula that uses $R_{ki}(\Theta^{[n]})$ and $\mathbf{W}^{[n+1]}$ to update $\beta_{ij}^{[n+1]}$:

$$\frac{1}{\beta_{ij}^{[n+1]}} = \sum_{k=1}^{K}R_{ki}(\Theta^{[n]})\left(Y_{ij} - \mathrm{CTF}_{ij}\sum_{m=1}^{M}\varphi_m(s_k)W_{mj}^{[n+1]}\right)^2. \tag{12}$$

During the iteration of the E-M algorithm, no class averages need to be computed. Nonetheless, upon convergence of the E-M algorithm, one can calculate the class averages through the posterior probability $R_{ki}$:

$$[A_k]_j = \frac{\sum_{i=1}^{N} R_{ki}(\Theta)\beta_{ij}\text{CTF}_{ij}Y_{ij}}{\sum_{i=1}^{N} R_{ki}(\Theta)\beta_{ij}\text{CTF}_{ij}^2 + \alpha} \quad . \tag{13}$$

Note that this equation is derived from Eqs. (11) and (12) and resembles the Wiener filter.

**2.3 Framework of unsupervised deep clustering**

Although SML-based algorithm may excel in unsupervised deep clustering, it is not computationally efficient in image alignment, a procedure that determines three geometrical parameters for each image: *x-y* translational shifts and in-plan rotation. Thus, we employed the adaptive MAP method (Scheres, 2012; Tagare, et al., 2010) to align single-particle images prior to SML-based clustering (Fig. 1b). Averages from random subsets of unaligned images are used to initialize the MAP-based image alignment. Further, a Gaussian model was used to initialize all the parameters of expectation-maximization algorithm in SML solution. Thus, in both the steps of MAP-based alignment and SML-based clustering, no external initial model or reference is used, ensuring the unsupervised nature of our approach.

Two strategies of deep clustering may be practiced. First, all particles are aligned based on the translations and rotations determined by MAP. Then SML is applied to partition these shifted and rotated particles into different subsets. CTF is corrected during class averaging similar to the Wiener filter. This first strategy is used for obtaining deep clustering quickly in several hours. However, some classes might be mixed with un-identical conformational states. Thus, in the second strategy, the dataset is initially classified into hundreds of classes, which are further classified into tens or hundreds of reference-free sub-classes per class, depending on the particle population of each class. Aligned particles of each subset are processed by SML classifier alone.

These two strategies can be used in different cases. If plenty of classes are needed to assess sample heterogeneity, the first strategy can be used to obtain relative more unsupervised classes without any human intervention. However, if more sub-classes are needed for some subsets, the second one can provide more details of specific subsets to examine structural heterogeneity. See Fig. 1b for flowcharts for these two strategies.

## 3 Results

**3.1 Classification accuracy on simulated data**

To quantify the classification accuracy of our approach, we generated a series of synthetic datasets with various SNRs, each of which is composed of 50,000 simulated images (see Supplementary Text 3 for details in image simulations). Since the original angles of the simulated data were known, we used the distribution of angular difference of an image pair classified into the same class to dictate the classification error. Smaller angular difference corresponds to smaller classification error or greater classification accuracy. With the dataset at a SNR of 0.02, the classification accuracy of our approach is almost the same as that of SPIDER and is better than EMAN2 and RELION (Supplementary Fig. 3c). For the data at a SNR of 0.01, our approach yielded the sharpest and highest peak at lower angular distances in the distribution curve (Fig. 2d) and obtained totally 495 effective classes (Fig. 2e). This feature remains when the SNR is further reduced to 0.005. By contrast, the maximum-a-posteriori (MAP) classification implemented in RELION obtained a relatively lower peak (Fig. 2d) but only 162 effective classes (Fig. 2e). Conventional PCA/K-means approaches in EMAN2 and SPIDER resulted in a much lower, wider peak at greater angular distances (Fig. 2d and Fig. 2e). When the SNR was decreased to 0.005, more prominent degrading behaviors were observed on the classification accuracy of these approaches (Fig. 2f, g and Supplementary Fig. 3e, f). These results suggest that the high level of noise significantly compromises the performance of data clustering by the conventional PCA/K-means clustering approach. In summary, our approach significantly improved the classification accuracy under conditions of reduced SNRs.

Furthermore, we examined the impact of dimensionality of the latent space on the classification accuracy in the SML algorithm. Using the same datasets of simulated images, we compared the measurements on the classification accuracy for the latent spaces assumed in one, two and three dimensions. The experiment with 2D latent space yielded slightly better classification accuracy than those with 1D and 3D latent space in our SML algorithm for SNR = 0.001, 0.0067 or 0.005 (Supplementary Fig. 2a, b, and Text 4). This result suggests that the increased dimensionality in manifold learning does not necessarily improve the classification accuracy.

## 3.2 Applications on experimental cryo-EM datasets

To demonstrate the applicability of our deep clustering approach, we applied our method to several experimental cryo-EM data (Supplementary Text 5 and Supplementary Fig. 1). First, ROME was used for an unsupervised clustering on a 17,103-particle dataset of the inflammasome complex (Zhang, et al., 2015). The resulting class averages directly reveal three different conformations of the complexes corresponding to 10-, 11- and 12-fold symmetry (Fig. 3a). For comparison, 300 classes were produced by 2D reference-free MAP classification implemented in RELION (Scheres, 2012) (Supplementary Fig. 4b). Second, unsupervised deep SML clustering of 96,488 particles of proteasomal RP-CP (regulatory particle associated with core particle) subcomplex directly reveals different local features among class averages (Chen, et al., 2016; Nogales, 2016) (Fig. 3b). These local differences were verified to correspond to distinct conformational states through further 3D reconstruction (Chen, et al., 2016; Nogales, 2016). Our results indicate that deep clustering can sort out not only the projections of underlying structure into relatively homogenous 2D classes but also unearth incomplete structures or junk particles, allowing efficient *in silico* purification of single-particle datasets.

Unsupervised clustering with improved accuracy can help identify hidden heterogeneity in the reference-free classes generated by other methods. For instance, after initial reference-free classification, particle images of specific classes could be selected for a deeper unsupervised clustering by SML. Delicate difference between SML-generated class averages may reveal features corresponding to distinct conformational states. To demonstrate this, a class of 281 particles resulting from MAP-based classification (Fig. 3c), whose average visually looks like the side view of 11-fold inflammasome complex, is further classified into 30 sub-classes by SML (Supplementary Fig. 5a). Based on the length of the intrinsic structure, the SML-based reference-free classification identified the side views of 10-fold and 12-fold inflammasome complexes among the 30 sub-classes (the red boxes in Fig. 3d), suggesting the MAP-generated initial class is mixed with the complexes with 10-fold, 11-fold and 12-fold symmetry. Another MAP-classified image group with 3961 particles, whose average visually looks like a tilted view of the free 19S regulatory particle (RP) that are not associated with a 20S core particle (CP), were further classified by SML (Fig. 3e). The deeper clustering identified 448 single-particle images (11.3%) of RP-CP subcomplexes that were mis-classified into the dataset of free RP (Fig. 3f). Comparison with other algorithms including MAP and K-means clustering suggests that our SML outperforms these algorithms in unambiguously identifying hidden heterogeneity in the initially classified data groups (Supplementary Text 6 and Supplementary Fig. 5).

### 3.3 Deep clustering improves initial reconstruction

To demonstrate deep clustering capacity of our approach, 117,471 particles of the free RP complex were classified into 1,000 classes by SML implemented in ROME in an unsupervised manner. The classification completed in 228 minutes with 30 rounds of E-M optimization on a computing cluster of 512 Intel Xeon E5 CPU cores (Supplementary Table 1) and 858 effective classes were obtained (Supplementary Text 7 and Supplementary Fig. 6). We used 234 class averages of the highest quality from the 1000 unsupervised classes to generate an initial 3D model of the RP complex in EMAN2 (Fig. 4a, Supplementary Fig. 6 and Supplementary Text 7). As a control, we used the conventional multivariate data analysis (MDA) and K-means method to obtain an initial model of RP complex in EMAN2 from 10,000 raw images of the RP complex (Fig. 4b and Supplementary Text 7). The FSC curves between the atomic model and the two initial models using 0.5 cutoff suggest that the resolution of the initial model using SML-based class averages was 20.6 Å, which is significantly higher that the resolution, 29.1 Å, of the other initial model obtained by the conventional approach (Fig. 4c). We conducted rigid-body fitting of the atomic model of the free RP into the two initial models (Fig. 4a and Fig. 4b). The contrasting result shows that the initial model built upon our deep clustering method is of significantly improved quality as opposed to that resulting from the conventional approach. Taken together, these results suggest that unsupervised deep clustering with both increased class number and improved classification accuracy in ROME gives rise to marked improvement in *ab initio* 3D reconstruction with an affordable computational cost.

### 3.4 Optimization of computational performance and efficiency

Our implementation of SML-based clustering algorithm in ROME software system has been optimized for modern CPU hardware, such as Intel® MIC Architecture used in the Intel Xeon Phi Knights Corner and Knight Landing processors. Our modernized code in ROME outperforms the existing software in both speed and magnitude. First, we compared the computational performance between ROME and RELION with several datasets of different particle numbers. It took 143 minutes for ROME to classify 57,001 particles of RP into 300 unsupervised classes on a computing cluster of 512 Intel Xeon E5 CPU cores (Fig. 5a). By contrast, it took 2379 minutes for RELION to classify the same dataset into 300 unsupervised classes on the same cluster. In all cases, the expectation-maximization was run for 30 rounds. Thus, under the same circumstance, ROME exhibits about 10-20 times speedup compared to RELION in unsupervised 2Dclassification. Second, a 96,488-particle dataset of RP complex was used to benchmark the performance of unsupervised clustering in ROME with different class numbers on the 512-core computing cluster (Fig. 5b). With increasing the class number from 100 to 1000, the running time of SML clustering is increased in a polynomial behavior from 14 to 214 minutes. Specifically, ROME completed unsupervised clustering of 1000 classes in 306 minutes with 30-round of expectation-maximization (Supplementary Table 1), whereas RELION won't complete the same rounds of optimization in a week on the same hardware condition. We also compared our MAP code with RELION under the same hardware environment (Supplementary Fig. 7). For more comparison, we also used this dataset to test reference-free clustering on 1000-class level by MSA and K-means clustering in SPIDER (Frank et al., 1996; Shaikh et al., 2008). Unfortunately, the program kernel was dumped upon a crash in the reference-free alignment step. More optimization details of ROME are provided in Supplementary Text 8 and 9. See Supplementary Table 1for more benchmark details of ROME.

## 4  Discussion

Unsupervised data clustering plays an important role in high-resolution structure determination by single-particle cryo-EM. In this study, we introduced SML-based deep clustering method following MAP-based image alignment, and implemented this algorithm in an open-source software ROME. One of the key findings of this study is that the SML-based clustering is more tolerant against noise than other approaches compared, including maximum-likelihood, PCA and K-means algorithms. This hallmark provides a path to an improvement in classification accuracy under lower SNR circumstances. Our SML algorithm built CTF correction into its mathematical kernel so that the classification won't be affected by the variation of defocus inherent in the cryo-EM data. Although the algorithm uses the E-M algorithm to maximize the regularized log-likelihood function in statistical nonlinear mapping between the latent and data spaces, which only guarantees the solution of a local optimum, the random initialization of the mapping ensures the unsupervised nature in data clustering. The class averaging expression under our SML scheme resembles the Wiener-type filter and gives rise to CTF-corrected, probability-weighted class averages, which inherits the advantage and traits of the maximum-likelihood based approaches (Scheres, 2012; Scheres, et al., 2005; Sigworth, 1998).

Utility strategies of deep clustering by SML can be highly versatile, problem-oriented, and user-controlled. In a typical scenario, one can pursue that all particles are aligned and partitioned into different unsupervised subsets in single runs of E-M optimization. This gives a glimpse of typical reference-free projection class averages of the protein and allows removal of junks. In other scenarios, particles of some specific classes can be taken out and further partitioned into deeper subsets. This may allow users to inspect the subtle differences inside the class to improve the structural homogeneity within classes. This application scenario can be interleaved with 3D classification to help examine the quality of 3D classes or clean up junks or poor quality images within 3D classes (Nogales, 2016; Zhang, et al., 2015).

In this work, we developed a statistical manifold learning based clustering approach for cryo-EM data processing. When optimized for Intel® Xeon® processors and Intel® Xeon Phi™ coprocessors, it exhibited an unprecedented capability in obtaining thousands of reference-free class averages in a highly efficient fashion. Importantly, a largely increased number of high-quality 2D class averages can lead to a more reliable initial reconstruction. Our applications to experimental datasets demonstrate the potential that unsupervised deep clustering by ROME is highly effective in distinguishing structural heterogeneity among single-particle datasets. This can in turn, when used iteratively with 3D classification, help in improving structural homogeneity of cryo-EM datasets, which has been already used for high-resolution refinement in our recent studies of several heterogeneous structure data (Chen, et al., 2016; Nogales, 2016; Zhang, et al., 2015).

**Acknowledgements**

The authors thank J. Jackson and T. Song for assistance in maintaining the high-performance computing system, K.M. Kermanshahche, D. Scott, C. Breshears, R. Ramanujam for helpful discussion. The experiments were performed in part at the Center for Nanoscale Systems at Harvard University, a member of the National Nanotechnology Coordinated Infrastructure Network (NNCI), which is supported by the National Science Foundation under NSF award no. 1541959. This work was funded by research funds at Peking University, by the Intel Parallel Computing Center program, and by gifts from Mr. and Mrs. Daniel J. Sullivan, Jr.

**Funding**

This work was funded by a grant of the Thousand Talents Plan of China (Y.M.), by a grant from National Natural Science Foundation of China 91530321 (Y.M.), by the Intel Parallel Computing Center program (Y.M.). The cryo-EM experiments were performed in part at the Center for Nanoscale Systems at Harvard University, a member of the National Nanotechnology Coordinated Infrastruc-ture Network (NNIN), which is supported by the National Science Foundation under NSF award no. 1541959.

Conflict of Interest: none declared.

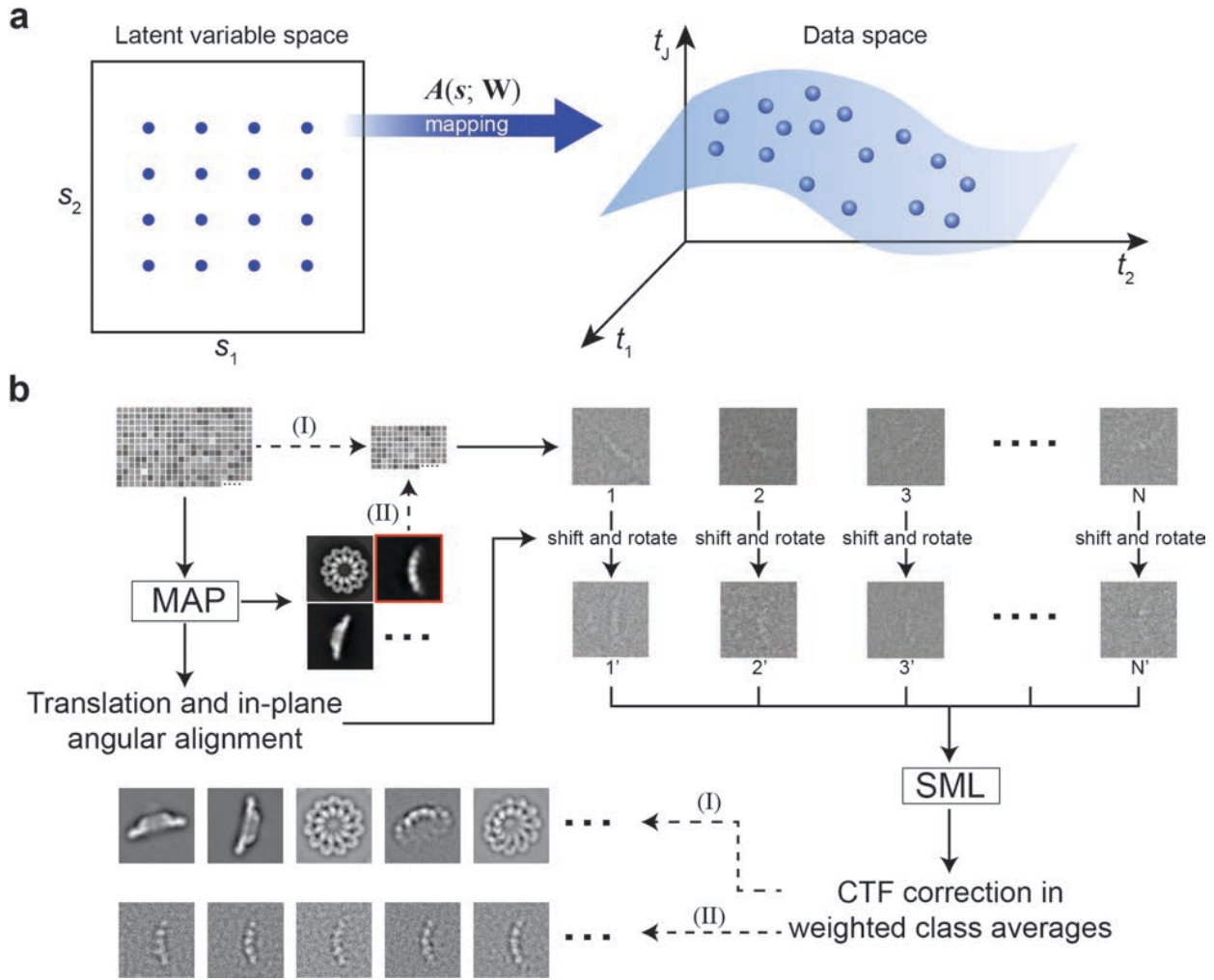

Fig. 1. Strategy of unsupervised deep clustering by statistical manifold learning. (a) The fundamental principle of SML is to establish the numerical relationship between variables in latent space and a non-Euclidean manifold composed of the Fourier transformed image data in D-dimensional data space. The manifold embed-ding can be determined by a set of nonlinear basis functions and a weighted parametric matrix. The likelihood function for the nonlinear mapping is solved by expectation-maximization method. (b) The work flow of two deep clustering strategies in ROME. (I) All images were aligned through MAP in a reference-free manner and then classified into a large number of groups by unsupervised SML. (II)

Unsupervised classes were further classified into many sub-classes by unsupervised SML. Initial supervised clustering can be done by either MAP or SML.

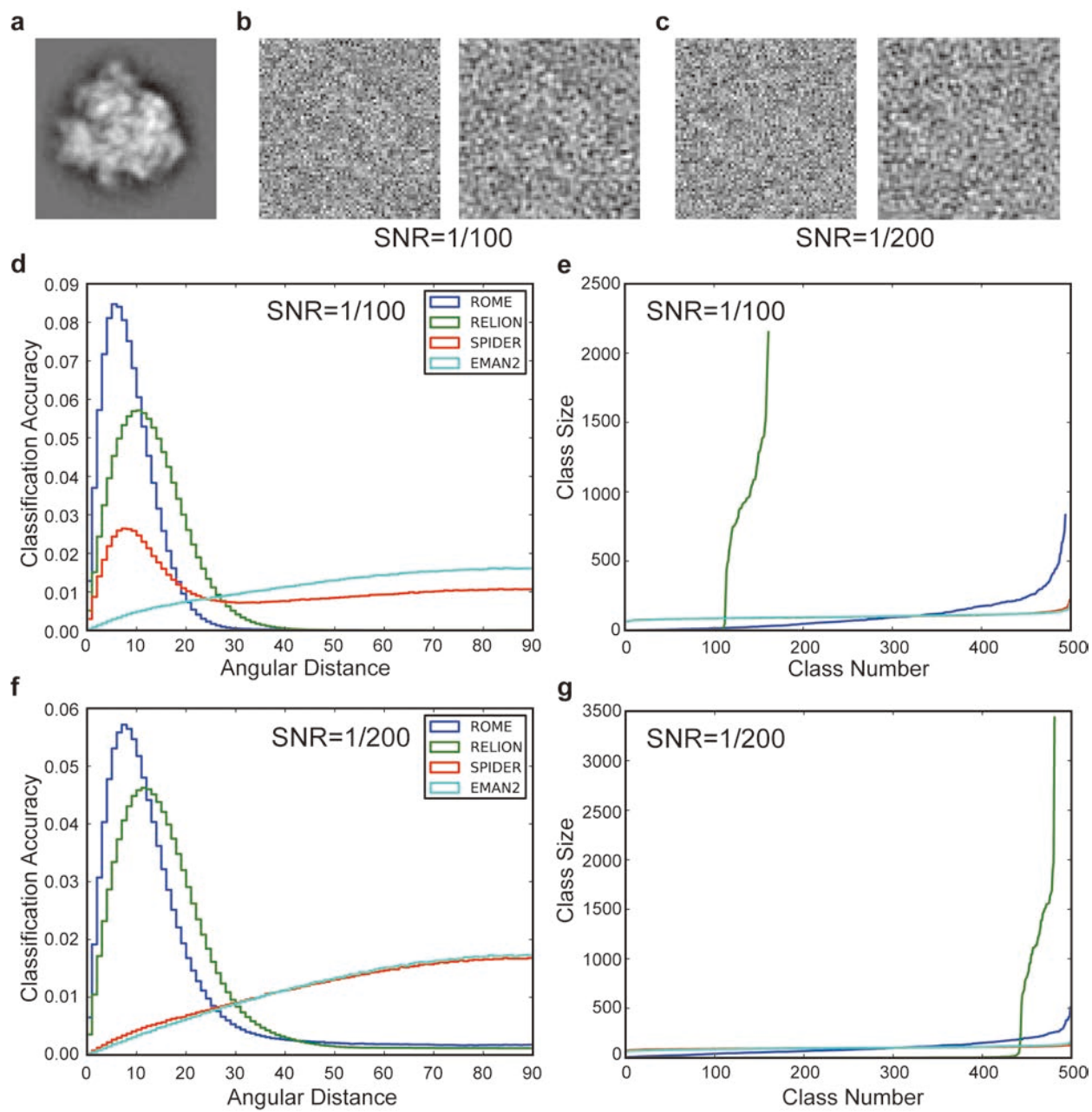

Fig. 2. Classification results of simulated data. (a) A projection of 70S ribosome model. (b) and (c) Examples of the corresponding simulated images of 70S ribosome with SNR = 1/100 (b) and SNR = 1/200 (c), respectively. The right panel in (b) and (c) shows the low-pass filtered version of each simulated image. (d) The normalized histogram exhibits the distributions of angular

distances resulting from the four classification methods that were applied to the simulated images of SNR=1/100. (e) The sizes of classes were ranked in ascend for the four classification methods when SNR=1/100. (f) Comparison of the distributions of angular distances resulting from four different classification methods that were applied to the simulated images of SNR=1/200. (g) Comparison of the class sizes ranked in ascend for the four classification methods when SNR=1/200.

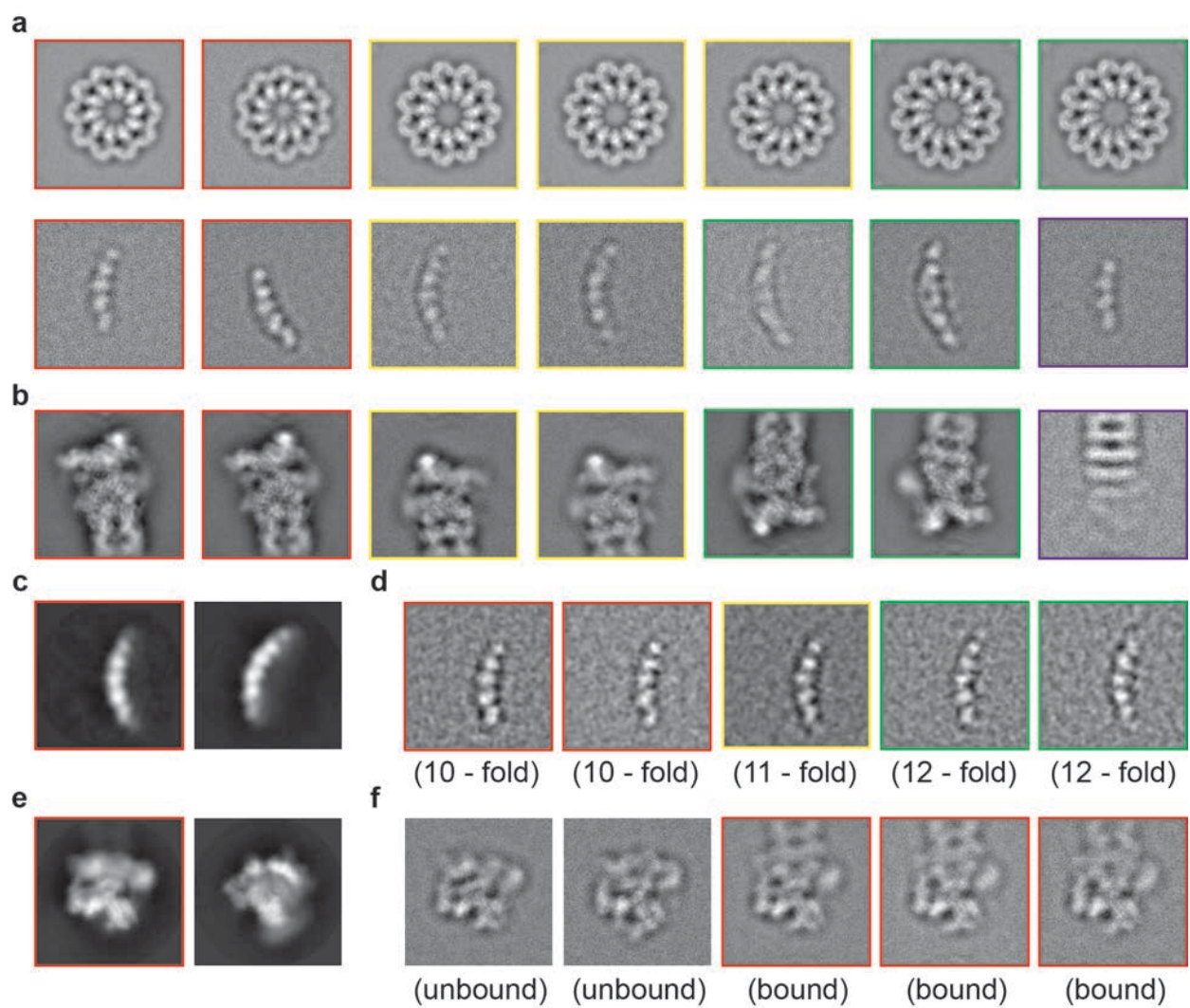

Fig. 3. Unsupervised clustering by SML. (a) A 17,103-particle dataset of inflammasome was classified with unsupervised SML in ROME. After alignment by MAP, three hundred classes were generated. Only top fourteen most particles class averages were shown. CTF-corrected

class averaging exhibit different views of inflammasome complex. In the first row, red, yellow and green boxes indicate the top views (the first row) and the side views (the second row) of 10-fold, 11-fold, and 12-fold inflammasome complex, respectively. The side views of the complex structure differ by length. Besides, the purple box exhibits a class average of an incomplete inflammasome complex. (b) A 96,488-particle dataset of RP-CP sub-complex was classified with unsupervised SML in ROME. Only top seven classes with the most particles among three hundred classes were shown. The red box or yellow box label a pair of class averages showing difference in local feature corresponding to the local movement of Rpn5 subunit of RP-CP subcomplex. The green box labels a pair of class averages showing the movement of Rpn1 subunit of RP-CP subcomplex. The purple box exhibits a class average of an incomplete RP-CP subcomplex. (c) A 16,306-particle dataset of inflammasome were initially classified with MAP classifier in a reference-free manner. Two classes among 50 classes visually looked like 11-fold inflammasome complex particles. (d) 281 particles of one class, whose average is highlighted by red box in panel c were further classified by SML. The red box exhibits the 11-fold inflammasome particles. The green box indicates the 10-fold inflammasome particles that are mis-classified by MAP into the same class as the rest 11-fold structures. The yellow box indicates the 12-fold inflammasome particles that are mis-classified by MAP into the same class as the rest 11-fold structures. (e) A 57,001-particle dataset of free RP were initially classified with MAP classifier in a reference-free manner. Most free RP complex and RP-CP subcomplex particles can be distinguished by 2D class averaging. Some mixed classes (red box) also exist. (f) 3,961 particles in one class in RP dataset (marked by red box in panel F) were further classified by SML in ROME. 448 RP-CP sub-complex particles were found to be misclassified into this free RP class.

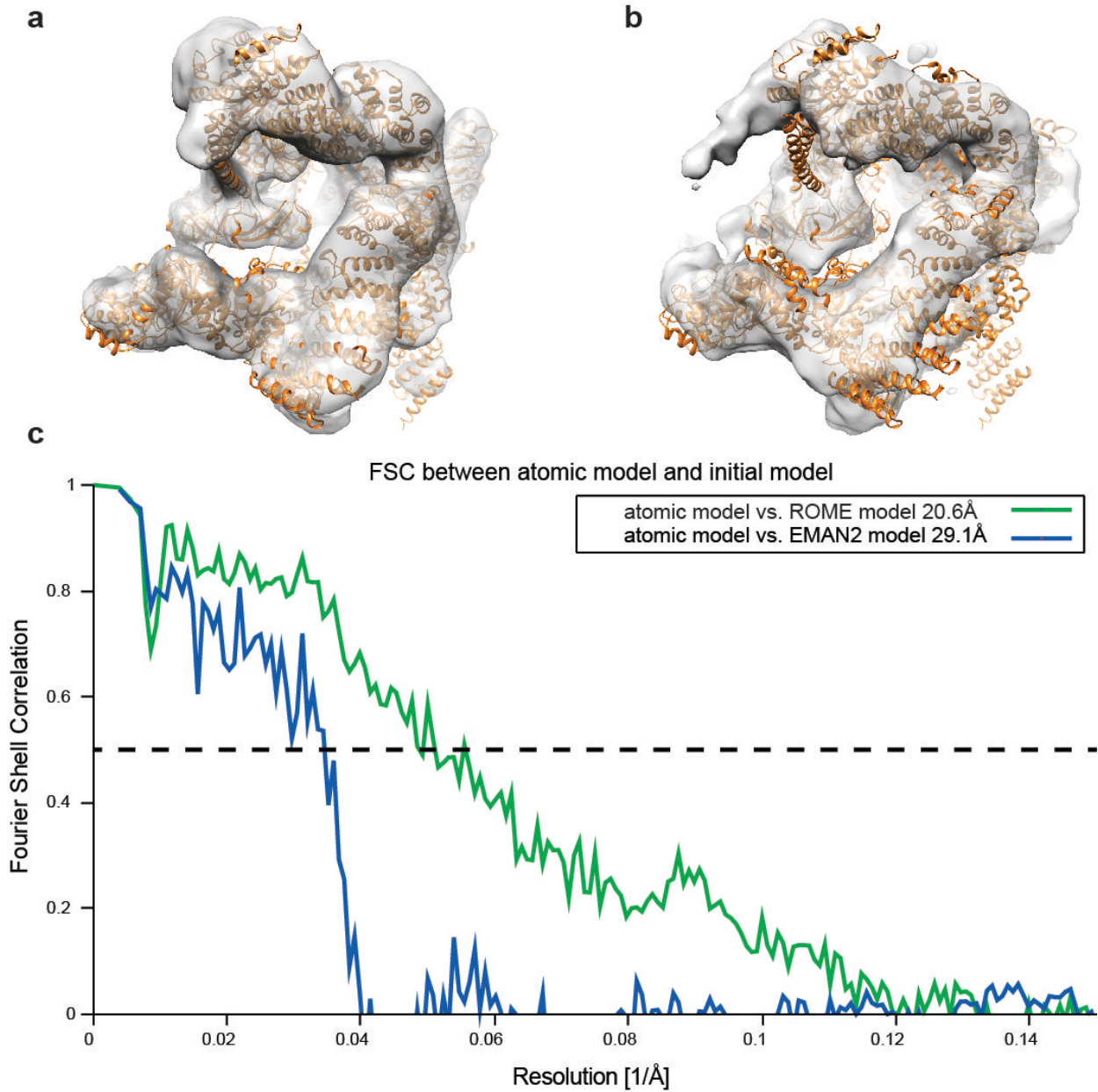

Fig. 4. Initial models reconstructed by the reference-free class averages of ROME and EMAN2. The RP atomic model was fitted in the density map, respectively. (a) The initial model calculated by the ROME-generated class averages is superimposed with the atomic model of free RP shown in a ribbon representation, suggesting that they are highly compatible with each other. (b) The initial model calculated by the EMAN2-generated class averages is superimposed with the atomic model of free RP shown in a ribbon representation. A substantial part of the atomic

model is out of the density of the initial model, suggesting poor map quality and a large shape error. (c) FSC between the RP atomic model and the initial models generated by ROME- and EMAN2-based class averages, respectively.

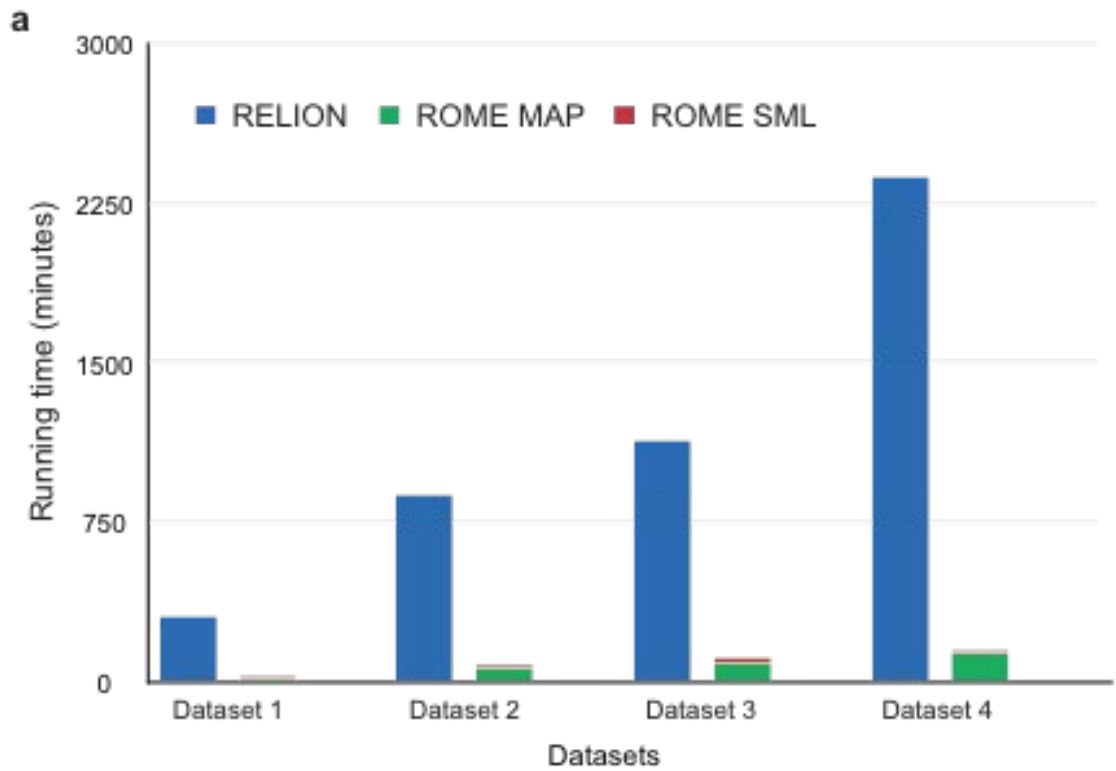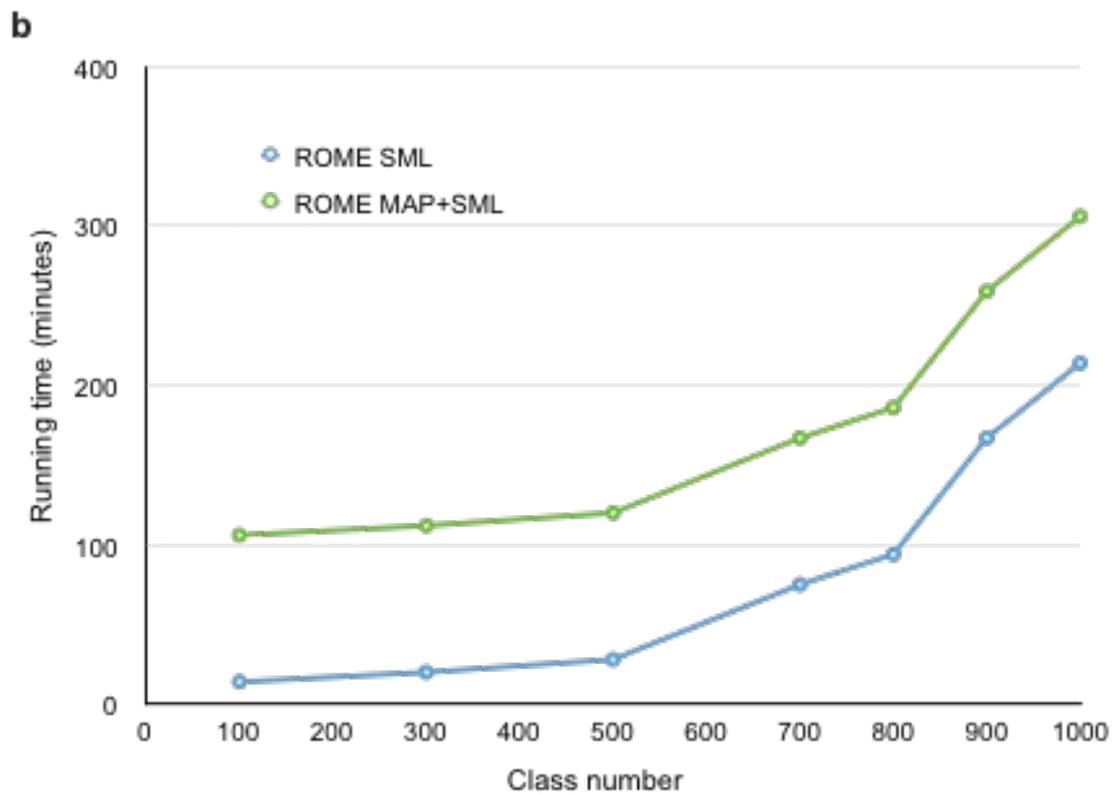

Fig. 5. Performance evaluation of unsupervised deep clustering with ROME and initial model generated by ROME class averages. (a) Performance of unsupervised clustering in ROME versus RELION through different data. Unsupervised 2D clustering into 300 classes based on RELION were performed on 4 experimental data (dataset1 denotes the 250*250 manually boxed 16306-particle dataset of inflammasome, dataset2 denotes the 160*160 manually boxed 35407-particle dataset of free RP complex, dataset3 denotes the 160*160 manually boxed 96488-particle dataset of RP-CP complex, dataset4 denotes the 160*160 manually boxed 57001-particle dataset of free RP complex), while MAP alignment in ROME and SML clustering for 300 classes were also performed. The blue histogram represents the running time of RELION. The green histogram represents the running time of MAP in ROME, and the red histogram represents the time of SML. (b) A 96,488-particle dataset of RP-CP subcomplex particles boxed by 180x180 were used to test the performance of SML in ROME (blue dots). Together with 92-minute alignment by MAP in ROME, the green dots represent the total running time including both alignment and clustering in ROME. The running time is polynomially related to class number.